\documentclass{article}

\usepackage{PRIMEarxiv}

\usepackage[utf8]{inputenc} 
\usepackage[T1]{fontenc}    
\usepackage{hyperref}       
\usepackage{url}            
\usepackage{booktabs}       
\usepackage{amsfonts}       
\usepackage{nicefrac}       
\usepackage{microtype}      
\usepackage{lipsum}
\usepackage{fancyhdr}       
\usepackage{graphicx}       
\graphicspath{{media/}}     
\usepackage{natbib}
\usepackage{amsmath}
\usepackage{multirow}
\usepackage{mathrsfs}
\usepackage[cal=cm,scr=boondox]{mathalpha}
\usepackage{mathtools}
\usepackage{makecell}

\title{A statistical perspective on transformers for small longitudinal cohort data}

\author{
  \parbox{0.95\linewidth}{\centering\bfseries
  Kiana Farhadyar$^{1,2}$\thanks{Corresponding author: Kiana Farhadyar (kiana.farhadyar@uniklinik-freiburg.de).},
  Maren Hackenberg$^{1,2}$,
  Kira Ahrens$^{3}$,
  Charlotte Schenk$^{3}$,
  Bianca Kollmann$^{4,5}$,
  Oliver T\"uscher$^{5,6,7,8}$,
  Klaus Lieb$^{5,6}$,
  Michael M. Plichta$^{3}$,
  Andreas Reif$^{3}$,
  Raffael Kalisch$^{5,9}$,
  Martin Wolkewitz$^{1,2}$,
  Moritz Hess$^{1,2}$,
  Harald Binder$^{1,2}$}\\[0.5em]
  \parbox{0.95\linewidth}{\centering\small
  $^{1}$ Institute of Medical Biometry and Statistics (IMBI), University of Freiburg, Freiburg, Germany \\
  $^{2}$ Freiburg Center for Data Analysis, Modeling and AI, University of Freiburg, Freiburg, Germany \\
  $^{3}$ Department of Psychiatry, Psychosomatic Medicine and Psychotherapy, University Hospital Frankfurt, Frankfurt, Germany \\
  $^{4}$ Department of Neuropsychology and Psychological Resilience Research, Central Institute of Mental Health (ZI), Mannheim, Germany \\
  $^{5}$ Leibniz Institute for Resilience Research (LIR), Mainz, Germany \\
  $^{6}$ Department of Psychiatry and Psychotherapy, Johannes Gutenberg University Medical Center Mainz, Mainz, Germany \\
  $^{7}$ Department of Psychiatry, Psychotherapy and Psychosomatic Medicine, University Medicine Halle, Martin-Luther University Halle-Wittenberg, Halle, Germany \\
  $^{8}$ German Center for Mental Health (DZPG), partner site Halle-Jena-Magdeburg, Halle, Germany \\
  $^{9}$ Neuroimaging Center (NIC), Focus Program Translational Neuroscience (FTN), Johannes Gutenberg University Medical Center, Mainz, Germany
  }
}

\begin{document}
\maketitle

\begin{abstract}
Modeling of longitudinal cohort data typically involves complex temporal dependencies between multiple variables. There, the transformer architecture, which has been highly successful in language and vision applications, allows us to account for the fact that the most recently observed time points in an individual's history may not always be the most important for the immediate future. This is achieved by assigning attention weights to observations of an individual based on a transformation of their values. One reason why these ideas have not yet been fully leveraged for longitudinal cohort data is that typically, large datasets are required. Therefore, we present a simplified transformer architecture that retains the core attention mechanism while reducing the number of parameters to be estimated, to be more suitable for small datasets with few time points. Guided by a statistical perspective on transformers, we use an autoregressive model as a starting point and incorporate attention as a kernel-based operation with temporal decay, where aggregation of multiple transformer heads, i.e. different candidate weighting schemes, is expressed as accumulating evidence on different types of underlying characteristics of individuals. This also enables a permutation-based statistical testing procedure for identifying contextual patterns. In a simulation study, the approach is shown to recover contextual dependencies even with a small number of individuals and time points. In an application to data from a resilience study, we identify temporal patterns in the dynamics of stress and mental health. This indicates that properly adapted transformers can not only achieve competitive predictive performance, but also uncover complex context dependencies in small data settings.
\end{abstract}


\section{Introduction}

Transformer neural network architectures \citep{vaswaniAttentionAllYou2017} have been successful for modeling of data with sequence structure, in particular for large language models \citep{radfordImprovingLanguageUnderstanding2018}. Key to this is the ability to take context effects into account, i.e. potentially complex patterns of elements that jointly occur in a sequence, when predicting the next element. Transformers have since been increasingly adopted for time series \citep{wenTransformersTimeSeries2023} and modeling of longitudinal data \citep{yangTransformEHRTransformerbasedEncoderdecoder2023, liBEHRTTransformerElectronic2020}. However, performance typically hinges on the availability of large amounts of training data \citep{hassaniEscapingBigData2022}. In small data settings, transformers might overfit and generalize poorly \citep{linSurveyTransformers2022}, and thus can be outperformed by much simpler models \citep{zengAreTransformersEffective2022}. 

To unlock the potential of transformers for longitudinal cohort data with few individuals and time points, we propose a minimal transformer model that relies on a considerably smaller number of parameters that need to be estimated. Instead of starting from the full transformer architecture and simplifying it, we use an established statistical model as a starting point, specifically a linear vector autoregressive model, and add in the minimally necessary ingredients of the multi-head attention mechanism that is at the core of the transformer architecture. 

Vector autoregressive (VAR) models, comprehensively covered by L\"utkepohl,\textsuperscript{\cite{lutkepohlNewIntroductionMultiple2005}}, have originally been designed to model multivariate time series by expressing the current time point as a linear combination of a fixed number of the most recent points, observed on an equidistant time grid. This does not allow for flexible-length patterns or a strong influence of non-recent time points. To allow for a larger number of characteristics, Bayesian VAR uses priors for regularization \citep{littermanForecastingBayesianVector1986}, and factor-augmented VAR projects variables into a lower-dimensional space \citep{bernankeMeasuringEffectsMonetary2005}. Several approaches have been developed to allow parameters or regimes to change over time \citep{tongThresholdAutoregressiveModelling1991, krolzigMarkovSwitchingVectorAutoregressive1997, tsayTestingModelingMultivariate1998}, some of which also allow for irregular sampling. Extensions such as those proposed by Cai et al.\textsuperscript{\cite{caiFunctionalCoefficientRegressionModels2000}} can model complex and nonlinear multivariate structures. Still, the number of past time points that are considered and needed for subsequent prediction has to be decided on beforehand. As an alternative, several artificial neural network architectures have been suggested, such as recurrent neural networks \citep{liptonCriticalReviewRecurrent2015} or LSTNet \citep{laiModelingLongShortTerm2018},  but still struggle  with long-range dependencies across the history of an observational unit.  

Transformer neural network architectures, which have been developed to address this,  can be seen as an extension of dynamic VARs \citep{luLinearTransformersVAR2025a} to deal with a varying number of past time points. Importance does not have to be inversely proportional in time, but is determined by an attention mechanism that assigns weights depending on a transformation of the characteristics of an observational unit observed at a time point. As a single attention pattern might be too limited \cite{cuiSuperiorityMultiHeadAttention2025}, transformers typically use multi-head attention, i.e. several weighting patterns. This can be interpreted as a nonparametric Bayesian ensemble, where each head approximates a sample from a posterior over attention distributions \citep{anRepulsiveAttentionRethinking2020}.  Yet, the interpretation of a fitted transformer model remains challenging. While attention scores have been proposed as indicators of feature relevance, their validity as explanations has been questioned, as attention weights may not reliably reflect the true importance of input features \citep{jainAttentionNotExplanation2019}. Recent work has focused on developing formal statistical frameworks for complex models \citep{colemanScalableEfficientHypothesis2022a, miwaStatisticalTestAnomaly2024}. Notably, \cite{shiraishiStatisticalTestAttention2025} introduced selective inference methods for vision transformers. However, these approaches do not specifically address the statistical significance of patterns in longitudinal cohort data, i.e. an approach that facilitates subsequent interpretation in this setting is missing so far. Enabled by the proposed minimal transformer architecture, we introduce a permutation-testing approach to fill this gap.

In Section \ref{sec:methods}, we introduce the minimal transformer architecture, MiniTransformer, also outlining  the main differences between the proposed architecture and the classical transformer, and providing a statistical testing framework. We evaluate our approach through a simulation design and an application to data from a cohort study on psychological resilience in Section~\ref{sec:experiments}. Section~\ref{sec:discussion} provides concluding remarks and a more general perspective on the potential of transformers for modeling small longitudinal data in biomedicine.

\section{Methods} \label{sec:methods}

\subsection{The MiniTransformer approach}

Consider the longitudinal data of an individual, where for each of $T$ successive time points $t_i, i=1,\ldots,T$, the vector $\mathbf{x}_{t_i} 
\in \mathbb{R}^{p+1}$ comprises $p$ measurements of characteristics of the individual and a constant term 1. The aim is to predict the values in $\mathbf{y} = (y_1, \ldots, y_{q})' $, observed at some future time $t_{T+1}$, which might be future measurements of the same variables, with $q=p$, or some other future characteristics. 

In a first step, we want to transform each $\mathbf{x}_{t_i}$ into scalar values $\tilde{x}^{(h)}_{t_i}, h=1,\ldots,H$, thereby implementing a lightweight version of multi-head attention to potentially reflect up to $H$ different patterns in the data. Specifically, the transformation of $\mathbf{x}_{t_i}$ should be able to take the information $\mathbf{x}_{t_l}, l=1,\ldots,T$, from another time point into account. To achieve this, we use a kernel function 
\begin{equation}\label{eq:kernel}
g(\mathbf{x}_{t_i}, \mathbf{x}_{t_l}; \mathbf{w}_{\text{query}}^{(h)}, \mathbf{w}_{\text{key}}^{(h)}) = \exp{\left(\mathbf{x}_{t_i}' \mathbf{w}_{\text{query}}^{(h)} \cdot \mathbf{x}_{t_l}' \mathbf{w}_{\text{key}}^{(h)}\right)} \cdot \exp{\left(-(w_{\mathrm{dist}}\cdot|t_i - t_l|)^\gamma\right)} 
\end{equation}
that encapsulates the idea of pairwise attention, parameterized by query parameters $w_{\mathrm{query}}^{(h)} \in \mathbb{R}^{p+1}$, key parameters $w_{\mathrm{key}}^{(h)} \in \mathbb{R}^{p + 1}$, and value parameters $\mathbf{w}_{\text{value}}^{(h)} \in \mathbb{R}^{p + 1}$, for each head $h=1,\ldots, H$. The second term in the product reflects the idea of decay of information as the temporal distance $|t_i - t_l|$ between the pair of observations increases, where $w_{\mathrm{dist}} \geq 0$ is a parameter that should be learned from the data, and $\gamma$ is a tuning parameter that we have set to $\gamma=5$ in our applications. 
Based on the kernel function (\ref{eq:kernel}), we obtain transformed values 
\begin{equation}
\label{eq:transform}
\tilde{x}_{t_i}^{(h)} = \frac{ \sum_{l=1}^i g(\mathbf{x}_{t_i}, \mathbf{x}_{t_l}; \mathbf{w}_{\mathrm{query}}^{(h)}, \mathbf{w}_{\mathrm{key}}^{(h)}) \cdot \mathbf{x}_{t_l}' \mathbf{w}_{\text{value}}^{(h)}}{\sum_{l=1}^i g(\mathbf{x}_{t_i}, \mathbf{x}_{t_l}; \mathbf{w}_{\mathrm{query}}^{(h)}, \mathbf{w}_{\mathrm{key}}^{(h)})} 
\qquad i=1,\ldots,T, \,\, h=1,\ldots,H
\end{equation}
that comprise weighted averages of the observations up to the current time point.

While $y \in \mathbb{R}^{q}$ could be predicted directly based on these transformed values, it will often be the case that there is some correlation structure between the elements, which can be explained by a lower-dimensional representation $z = (z_1, \ldots, z_C)$ with $C < q$. Therefore, we suggest to cumulate the transformed values $\tilde{x}_{t_i}^{(h)}$ into
\begin{equation}\label{eq:cumulant}
z_c = \sum_{i=1}^T \tilde{\mathbf{x}}_{t_i}' \mathbf{w}_{\mathrm{cum}}^{(c)} \cdot \exp{\left(-(w_{\mathrm{horizon}}\cdot|t_{T+1} - t_i|)^\gamma\right)} \quad c=1,\dots,C,
\end{equation}
where each cumulant $z_c$ can receive input from the $H$ different heads via the vector of transformed values $\tilde{\mathbf{x}}_{t_i} = (\tilde{x}_{t_i}^{(1)}, \ldots, \tilde{x}_{t_i}^{(H)})' $, as parameterized via $\mathbf{w}_{\text{cum}}^{(c)} = (w_{\mathrm{cum},1}^{(c)}, \ldots, w_{\mathrm{cum}, H}^{(c)})' $. There, each contribution receives a decay according to the distance $|t_{T+1}-t_i|$ to the prediction horizon time $t_{T+1}$, at which $\mathbf{y}$ is observed, parameterized by $w_{\mathrm{horizon}} \geq 0$.

Predictions are then obtained via a standard regression model
\begin{equation}
\label{eq:output}
\hat{y}_r = \beta_0^{(r)} + \mathbf{z}'\boldsymbol\beta^{(r)}, \quad r = 1,\ldots,q,
\end{equation}
with intercept parameters $\beta_0^{(r)}$ and parameter vectors $\boldsymbol{\beta}^{(r)} = (\beta_1^{(r)}, \ldots, \beta_{C}^{(r)})', c=1,\ldots,C,$ for transforming the input of the cumulant vector $\mathbf{z}=(z_1,\ldots,z_C)'$.

Estimates for all parameters can be obtained jointly, by minimizing the squared error loss $\sum_{r=1}^{p} (y_r - \hat{y}_r)^2$. If other types of endpoints are to be considered, e.g., binary or time-to-event endpoints, appropriate alternative regression models and loss functions can be used instead. Assuming that data for several individuals is available, batched stochastic gradient descent \citep{bottouLargeScaleMachineLearning2010} is used for parameter estimation, with gradients obtained via differential programming techniques \citep{hackenbergUsingDifferentiableProgramming2020}.

\subsection{Main differences to standard transformers}
\label{sec:minitransformer_differences}

In a standard decoder-based transformer architecture \citep{vaswaniAttentionAllYou2017}, each element of a sequence, typically called ``token'' is mapped into a high-dimensional space via a neural network embedding layer. In text processing applications, where the tokens are (parts of) words, this is necessary to obtain a quantitative representation for subsequent modeling. Such an embedding layer could also be considered in longitudinal cohort settings, where the tokens, i.e., the observations in the history of an individual, are already quantitative, to increase representation capacity. However, such a transformation could make interpretation of the effect of individual characteristics considerably more difficult, particularly in small data settings where overparameterization is also a concern \citep{barceloModelInterpretabilityLens2020}. Therefore, the MiniTransformer approach omits this embedding and just adds a constant element to the input vectors to allow for shifts in representation.

As a further simplification, the query and key projections in (\ref{eq:kernel}) and the value projection in (\ref{eq:transform}) reduce the representations to scalars, with weight parameter vectors $\mathbf{w}_{\mathrm{query}}^{(h)}$, $\mathbf{w}_{\mathrm{key}}^{(h)}$, and $\mathbf{w}_{\mathrm{value}}^{(h)}$, for each attention head $h=1,\ldots,H$. In contrast, standard transformers rely on weight matrices to project into query, key, and value representations, typically of the same dimension $q_{\mathrm{QKV}} \geq p$. While the MiniTransformer does not have that much representational flexibility, some of this might be recovered by still allowing for the input of multiple attention heads in (\ref{eq:cumulant}), to potentially reflect different types of patterns in the data. 

In standard transformers, the information about the order of observations is incorporated through either absolute positional encoding \citep{vaswaniAttentionAllYou2017} or relative positional encoding \citep{shawSelfAttentionRelativePosition2018}, each with flexible influence for each index position. This is combined with a clipping value $\mathscr{c}$, beyond which all distances are treated equally, resulting in at least $\mathscr{c} + 1$ parameters that need to be estimated. The MiniTransformer uses a relative positional encoding schemes in (\ref{eq:kernel}) and (\ref{eq:cumulant}), but reduces complexity to just three parameters $w_{\mathrm{dist}}$, $w_{\mathrm{horizon}}$, and $\gamma$. This still allows for reflecting patterns where the influence of observations smoothly diminishes over time (e.g. \citealp{setodjiExponentialEffectPersistence2019,schillingEmotionalReactivityDaily2022, liGenericMedicalConcept2022}).

Finally, the MiniTransformer simplifies the standard practice of transforming the concatenated multi-head output, in our case $\tilde{\mathbf{x}}_{t_i}$, by a large projection layer. Instead of implementing the latter as a fully connected neural network (with $\gg (q_{QKV}\cdot H)^2$ parameters), we just use a linear transformation via \(\mathbf{w}^{(.)}_\mathrm{cum}\) to summarize information across attention heads. While the further linear transformation in (\ref{eq:output}) might seem redundant, we found in our experiments that this provided more flexibility when combined with random parameter initialization and stochastic gradient descent. Non-linear transformations, as often employed in standard transformers, could be easily introduced in (\ref{eq:transform}), (\ref{eq:cumulant}), and (\ref{eq:output}) to ease identifiability concerns, but we found no systematic benefit in our experiments.

Taken together, the simplifications in the MiniTransformer reduce the number of parameters to be estimated by at least $\mathcal{O}(p^2H^2)$, while still preserving core ideas of attention and temporal decay.

\subsection{A permutation test for context effects}\label{sec:statistical_testing}

Based on the attention kernel (\ref{eq:kernel}), the MiniTransformer approach determines the influence of each of the previous observations, which serve as a context to the current observation, on the transformation of the current observation in (\ref{eq:transform}), and thus also its contribution to the subsequent prediction in (\ref{eq:output}). Therefore, the estimated parameters reflect context patterns, and in particular, which characteristics of individuals contribute the most to such context patterns. Unfortunately, the individual parameters are difficult to interpret, and also not amenable to statistical inference. Therefore, we focus on the effect of characteristics of individuals on the context that are reflected in changes in prediction. For this, we propose
a permutation testing approach that compares predictions that are based on a single observation without context to predictions with context observations. 

Specifically, we consider just a single previous observation $\mathbf{x}^{\mathrm{context}} = \left(1, x^\mathrm{context}_{1}, \ldots, x^\mathrm{context}_{p}\right)'$ as a potential context for each of $V$ potential later observations $\mathbf{x}^{\mathrm{visit}}_v = \left(1, x^\mathrm{visit}_{v,1}, \ldots, x^\mathrm{visit}_{v,p}\right)', v=1,\ldots,V $, that might subsequently be used for predicting a future value $y_r$. We will ignore precise time information (essentially setting the time decay factors in (\ref{eq:kernel}) and (\ref{eq:cumulant}) to the value $1$). Therefore, we will also omit time indices. 

The effect of variable $j\in\{1,\ldots,p\}$ on the difference between the contribution of $\mathbf{x}_v^{\mathrm{visit}}$ to the prediction of $\hat{y}_r$ with and without context is
\begin{equation}\label{eq:diffpred}
\Delta^{(r)}_{j, v} =
\sum_{c=1}^{C} \beta_{c}^{(r)} \cdot \sum_{h=1}^{H}  w_{\mathrm{cum}, h}^{(h)} \cdot \frac{
  \sum_{j=1}^{p} w^{(h)}_{\textrm{value},j}\cdot \left(x_{v,j}^\mathrm{visit} - \mathbf{x}_{j}^\mathrm{context}\right)}
{1 +g(\mathbf{x}_v^{\mathrm{visit}}, \mathbf{x}_v^{\mathrm{visit}}; .)/g(\mathbf{x}_v^{\mathrm{visit}}, \mathbf{x}^{\mathrm{context}}; .)},
\end{equation}
which in particular depends on how $\mathbf{w}_{\textrm{value}}^{(h)} = (w_{\textrm{value}, 0}^{(h)},w_{\textrm{value}, 1}^{(h)}, \ldots, w_{\textrm{value}, p}^{(h)})'$ weights the differences between the variable with index $j$ in $\mathbf{x}_v^{\mathrm{visit}}$ and $\mathbf{x}^{\mathrm{context}}$. The difference (\ref{eq:diffpred}) also increases when the attention kernel $g(\cdot)$ between the two observations becomes larger relative to the self-attention of $\mathbf{x}_v^{\mathrm{visit}}$, which can also be driven by the variable with index $j$.

Assuming that the variables in $\mathbf{x}^{\mathrm{context}}$ have similar variance, which could be facilitated by standardization before model fitting, we can now add some value $\delta$ either to the variable with index $j_1$ in $x^{\mathrm{context}}$, or to the variable with index $j_2$, to induce a comparable change, and compare the two changes by calculating (\ref{eq:diffpred}) for each of the two, to obtain $\Delta_{j_1,v}^{(r)}$ and $\Delta_{j_2, v}^{(r)}$. More generally, this could be performed for all variables, and all visit observations ${\mathbf{x}^{\mathrm{visit}}_{v}}$, resulting in a matrix
\begin{equation}\label{eq:diffmat}
\boldsymbol{\Delta}^{(r)} := \left( \Delta^{(r)}_{j, v} \right)_{\substack{j=1, \ldots, p \\ v=1, \ldots, V}}.
\end{equation}
This matrix depends on the values of $\mathbf{x}^{\mathrm{context}}$ and $\delta$, and these should be chosen so that $\mathbf{x}^{\mathrm{context}}$ represents some kind of reference observation, and $\delta$ a change of reasonable magnitude. For example, when there is only binary data with values $0$ and $1$, all values of $x^{\mathrm{context}}$ might be set to zero, and $\delta$ set to $1$.

Summary statistics $s_j^{(r)}, j=1,\ldots,p,$ can now be calculated across the rows of $\boldsymbol{\Delta}^{(r)}$ to assess the impact of each variable as part of the context. We suggest considering the average row values to capture directional effects, but the average of squared values, or something else, might also be useful. 

Under the null hypothesis of no distinct context effect of the variable with index $j$, the rows of $\boldsymbol{\Delta}^{(r)}$ should be exchangeable. Therefore, we calculate summary statistics $s_j^{(r,m)}$ for $m=1,\ldots,M$ permutations, and obtain a p-value
\begin{equation}
\text{pval} = \frac{1}{M}
\sum_{m=1}^{M} I\left(s_j^{(r)} \geq s_j^{(r,m)}\right)
\label{eq:p-value}
\end{equation}
that reflects the context effect on $y_r$, where \(I(\cdot)\) is the indicator function that returns one if the condition is true and zero otherwise. The empirical null distribution may be contaminated by other variables that play a significant role in the context, but this will only lead to a conservative behavior of (\ref{eq:p-value}).

While we have focused on a single prediction target $y_r$ so far, we can consider all elements of $\mathbf{y}$ to obtain the matrix
\(\mathbf{S} =
\left( s_{j}^{(r)} \right)
\substack{j = 1, \ldots, p \\ r = 1, \ldots, q}\), e.g. for visualization as a heatmap.

\section{Empirical evaluation}\label{sec:experiments}

We evaluated the MiniTransformer both with simulated and real data. To ensure reproducibility, we provide an implementation of the MiniTransformer architecture, the permutation testing approach, and all evaluation procedures at \url{https://github.com/kianaf/MiniTransformer}. This repository includes separate Jupyter notebooks for each experiment, documenting all data preprocessing steps, hyperparameters, and random seeds.

\subsection{Simulation study}\label{sec:simulation_experiments}

We generate sequences of binary data $\mathbf{x}_{t_i} \in \{0,1\}^{p}, i=1,\ldots,T$. Where the value of the variable with index $j_3$ depends on past values of the variables with indices $j_1$, $j_2$ and $j_3$. This dependency structure is governed by an unobserved variable $z_{t_i}$, which encodes whether a specific pattern of activation has occurred and persisted up to time $t_i$. Specifically, $z_{t_i}$ is set to 1 if there exists a pair of earlier time points $i_1 < i_2 \leq i$ such that variable $j_1$ was active at time $t_{i_1}$, variable $j_2$ was active at time $t_{i_2}$, and variable $j_3$ has remained inactive from time $t_{i_2}$ onwards. Otherwise, $z_{t_i} = 0$. This condition can be formally written as:

$$
z_{t_i}
=
\begin{cases}
1, &
\text{if } \exists\, i_1<i_2 \leq i 
\text{ such that }  x_{t_{i_1},j_1} = x_{t_{i_2},j_2} = 1
\;\land\;
\forall i^* > i_2:\; x_{t_{i^*},j_3} = 0, \\
0, & \text{otherwise.}
\end{cases}
$$
The observed values are iteratively generated as
$$
x_{t_{i+1},j} = 
\begin{cases}
I\left(
z_{t_i} = 1
\;\land\;
B_{t_i} = 1,\;
B_{t_i} \sim \operatorname{Bernoulli}(0.9)
\right)
& j = j_3,\\
\mathrm{Bernoulli}(0.7), &  j \neq j_3,\\
\end{cases}
$$
and for $i>2$ the sequence is terminated with probability 0.2 to obtain sequences of different length, with at least 3 observations.

To assess the impact of the number of sequences on performance, we trained the MiniTransformer on datasets comprising 100, 500, or 1000 sequences, each with $p=10$ variables. For estimating the parameters, the task is to predict all $p$ variables of the last observation of a sequence, and the same for all subsequences of at least length 3 that can be obtained by starting at the first time point and gradually adding more time points. We used an architecture with $H = 12$ heads and $C=2$ cumulants, and estimated parameters via stochastic gradient with batches of size 1 and a learning rate of 0.001 for 100 epochs.

For comparison, we consider three types of approaches: The first approach (Average) uses the average of all observations of a variable on the training data to predict on the test data. As a second approach (Regression), we consider a regression model for each variable that uses the values of all variables from the previous time point to predict the next value of that specific variable. Finally, we consider an approach  (Informed) that is partially informed by the true underlying structure. For all variables except $x_{t_i,j_3}$, it uses the average for prediction, just as the first approach. For $x_{t_i,j_3}$, it calculates the average of all values conditional on $x_{t_i,j_2}$ being 0 or 1 at the previous time point, and subsequently uses this for prediction.

\begin{table}[tb]
    \centering
    \footnotesize
    \caption{Prediction performance of the MiniTransformer for simulated data with $p=10$ variables in training datasets with different numbers of sequences $n_\text{train}$, compared to a simple averaging approach, regression models, and an approach informed by the true structure. Mean squared error is considered across all variables (MSE) and specifically for the variable that is affected by the true pattern (MSE$_{j_3}$), together with standard deviations. Smallest values are indicated by boldface.}
    \label{tab:sim}
\begin{tabular}{llcccc} 
        \toprule
        \textbf{Approach} & \textbf{Metric} & $n_\text{train} = 100 $ & $n_\text{train} = 200$ & $n_\text{train} = 500$ & $n_\text{train} = 1000$ \\
        & & & & & \\ 
        \midrule
\multirow{2}{*}{\makecell{Average}} & MSE & $0.209 \pm 0.002$ & $0.21 \pm 0.002$ & $0.209 \pm 0.002$ & $0.209 \pm 0.002$ \\
 & MSE$_{j_3}$ & $0.218 \pm 0.007$ & $0.217 \pm 0.009$ & $0.214 \pm 0.01$ & $0.211 \pm 0.007$ \\
\addlinespace
\multirow{2}{*}{Regression} & MSE & $0.209 \pm 0.003$ & $0.206 \pm 0.002$ & $0.204 \pm 0.002$ & $0.204 \pm 0.002$ \\
 & MSE$_{j_3}$ & $0.155 \pm 0.005$ & $0.153 \pm 0.006$ & $0.152 \pm 0.005$ & $0.151 \pm 0.004$ \\
\addlinespace
\multirow{2}{*}{Informed} & MSE & $\mathbf{0.204 \pm 0.002}$ & $0.204 \pm 0.002$ & $0.204 \pm 0.002$ & $0.204 \pm 0.002$ \\
 & MSE$_{j_3}$ & $0.163 \pm 0.008$ & $0.163 \pm 0.008$ & $0.161 \pm 0.008$ & $0.159 \pm 0.005$ \\
\addlinespace
\multirow{2}{*}{MiniTransformer} & MSE & $0.21 \pm 0.006$ & $\mathbf{0.202 \pm 0.002}$ & $\mathbf{0.196 \pm 0.001}$ & $\mathbf{0.195 \pm 0.002}$ \\
 & MSE$_{j_3}$ & $\mathbf{0.141 \pm 0.03}$ & $\mathbf{0.097 \pm 0.013}$ & $\mathbf{0.068 \pm 0.006}$ & $\mathbf{0.06 \pm 0.007}$ \\
        \bottomrule
    \end{tabular}
\end{table}

For evaluation, we consider the mean squared error (MSE) of prediction for 10 repetitions on a test dataset with 1000 sequences. Specifically, we consider the average MSE when predicting all $p$ variables at the last time point of each sequence, and also the MSE just for $x_{t_i,j_3}$. As seen from Table \ref{tab:sim}, the MiniTransformer is competitive across different training sample sizes, even with just 100 sequences. It is only outperformed by the informed approach for the smallest sample size when considering the MSE across all variables, but not when considering the MSE just for $x_{t_i,j_3}$. This might indicate some overfitting for the other variables by the MiniTransformer due to the small sample size.

Next, we assess whether the learned cumulants recover the known underlying structure \(z_t\) in the simulation. As shown above, when $z_t=1$, the probability of the target $x_{t,j_3}$ increases sharply, so recovering $z_t$ provides a direct test of whether the learned representation captures the intended mechanism rather than merely improving prediction.

\begin{figure}[htbp]
    \centering
    \includegraphics[width=0.9\linewidth]{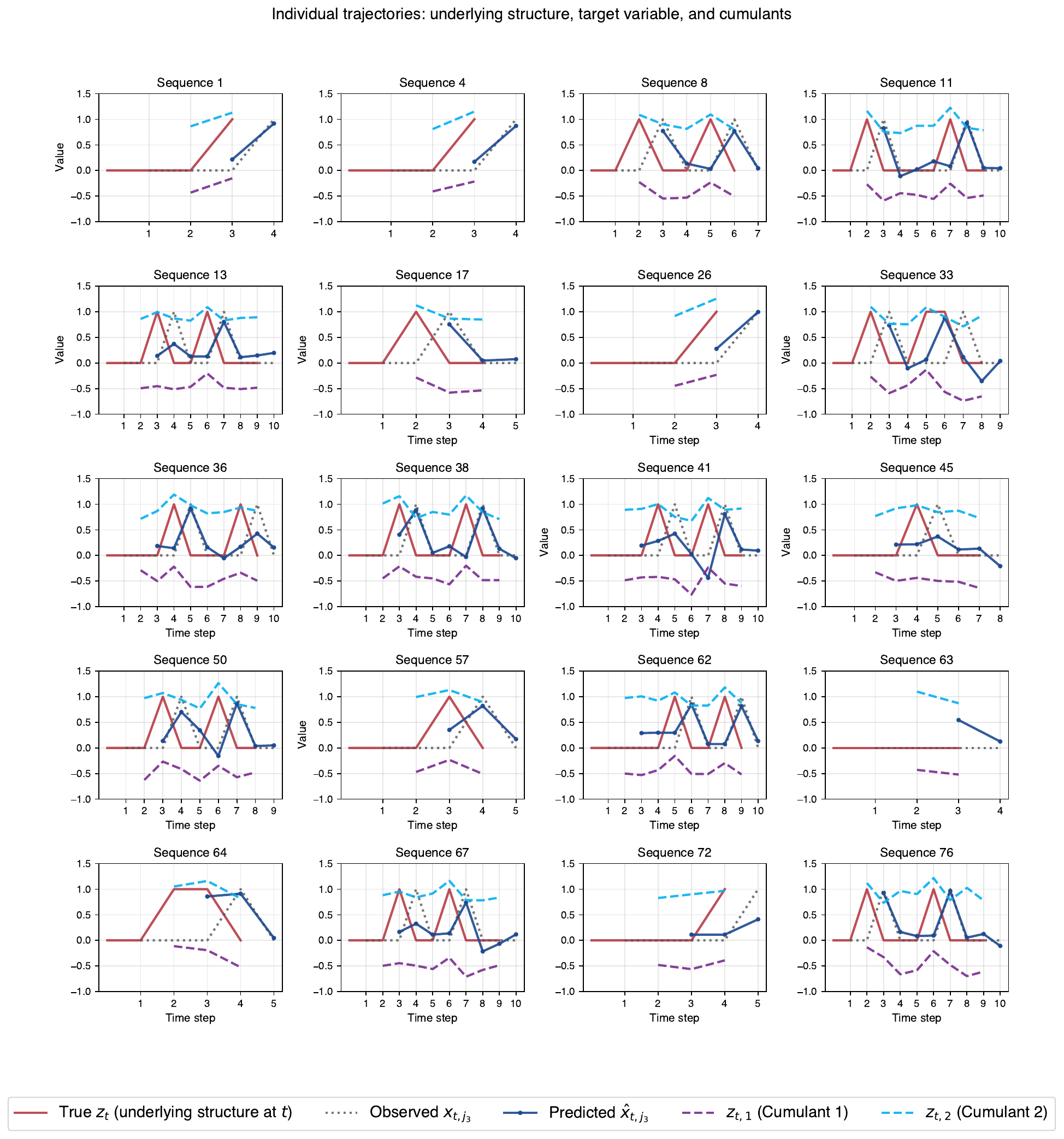}
    \caption{Individual simulated and fitted trajectories for a random selection of individuals, with true underlying structure $z_t$, observed target $x_{t,j_3}$, predicted target $\hat{x}_{t,j_3}$, and learned cumulants (~\ref{eq:cumulant}). 
    }
    \label{fig:cumulant_trajectories}
\end{figure}

Figure~\ref{fig:cumulant_trajectories} shows a random subset of simulated sequence trajectories, overlaying the true underlying structure $z_t$, the observed and predicted value of \(x_{t,j_3}\), and the learned cumulants (\ref{eq:cumulant}).
Across sequences, the cumulant trajectories rise and fall in synchrony with the onset/offset of $z_t$, indicating that the cumulant representation captures the latent mechanism.
Moreover, predicted target trajectories align with the observed target in periods where the latent structure is active, consistent with the simulation design in which $z_t$ modulates the target dynamics.

To assess the statistical testing approach, we consider the most challenging setting with just 100 sequences, and the variable with index $j_3$ as the target for prediction. As a full permutation approach with all potential realizations of $\mathbf{x}_v^{\mathrm{visit}}$ is not feasible, we randomly sample 8 patterns. Across the 10 repetitions, the variable with index $j_1$ then receives an average p-value of $0.146$, with index $j_2$ an average of $0.1285$, and with index $j_3$ an average of $0.0022$. The other seven variables receive averages between $0.2242$ and $0.6777$. This indicates that the approach can uncover the variables involved in the true underlying pattern, and in particular the variables with indices $j_1$ and $j_2$ as the driving factors of context effects.

We also wanted to evaluate how well the sampling approach for realizations of $\mathbf{x}_v^{\mathrm{visit}}$ reflects what would be obtained with a full permutation approach. For computational feasibility, we consider a setting with just $p=4$ variables, three of which carry the true pattern, and a total of 50 sequences. 

The sampling-based p-values for $V=8$ and $V=12$ realizations of $\mathbf{x}_v^{\mathrm{visit}}$ with 10 repetitions are contrasted with the p-values based on the full permutation approach in Table~\ref{tab:statistical_testing_results_sim}. Similar to the results for 100 or more sequences, the three variables that are part of the underlying true pattern receive the smallest p-values. Yet, the sampling-based approach seems to have less power and more variability, in particular for a smaller number $V$ of visit observation samples. Still, it is surprising that even with just 50 sequences, some parts of the true pattern can still be identified.

\begin{table}[tbp]

    \centering
    \footnotesize
    \caption{Average p-values (and standard deviations from 10 repetitions) resulting from the permutation testing approach when using sampling, with either $V=8$ or $V=12$ patterns of $\mathbf{x}_v^{\mathrm{visit}}$ sampled, contrasted with the full permutation approach with $V=16$, in a setting with $p=4$ variables (where variables 1-3 are part of the true pattern) and 50 sequences.}
    \label{tab:statistical_testing_results_sim}
     \begin{tabular}{cccc}
        \toprule
        Variable & 
        $V=8$ & 
        $V=12$ & 
        $V=16$\\  
        \midrule
         1  & \(0.084 \pm 0.13\)   & \(0.01 \pm 0.02\)   & \(<0.001\)  
            \\
         2  & \(0.2 \pm 0.29\)   & \(0.049 \pm 0.06\)   & \(0.002\)  
            \\
         3  & \(0.084 \pm 0.16\)   & \(0.01 \pm 0.02\)   & \(0.007\)  
            \\
         4  & \(0.607 \pm 0.24\)   & \(0.633 \pm 0.18\)   & \(0.635\)  
            \\
        \bottomrule
    \end{tabular}
\end{table}

\subsection{Real Data: reported stressors and general mental health status}
 We consider the longitudinal resilience assessment (LORA) study \citep{chmitorzLongitudinalDeterminationResilience2021} as an application that motivated our development of the MiniTransformer approach. Specifically, we focus on three questionnaires that the participants answered every three months: the Mainz Inventory of Microstressors (MIMIS), a 58-item instrument capturing self-reported daily hassles (microstressors, dh) in the past week \citep{chmitorzAssessmentMicrostressorsAdults2020}; a 27-item life events inventory assessing self-reported life events (macrostressors, le) over the past three months \citep{canliNeuralCorrelatesEpigenesis2006}; and the 28-item General Health Questionnaire (GHQ-28), which measures mental health status over the past weeks \citep{goldbergValidityTwoVersions1997}. The GHQ-28 consists of four subscales, including somatic symptoms, anxiety/insomnia, social dysfunction, and depression, each containing seven items. Participants reported the number of days they experienced each daily hassle in the past week and the number of major life events that occurred in the past three months. The GHQ-28 items were rated on a 4-point Likert scale ranging from 0 (“Not at all”) to 3 (“Much more than usual”), with total and subscale scores calculated by summing the respective item ratings. 
 
 To illustrate our approach, we consider two different subsets of items, each with \(p = 10\) variables. In both datasets, we have one variable related to the general mental health status as a target variable. The other variables are stressors selected from macro- and microstressors, which we want to investigate for their temporal interaction with the target. Items were randomly selected from the available MIMIS and life events inventories, resulting in a mix of both microstressors (dh) and macrostressors (le) without pre-filtering for theoretical relevance to the target. This random selection allows us to assess whether the statistical testing framework can identify context effects in an unbiased manner. We preprocessed the data to obtain binary variables. Specifically, for the stressors, if a participant reports the presence of an item, we record this as 1, otherwise as 0. The GHQ total is dichotomized with a threshold of (\(>23\)) and GHQ subscales with thresholds (\(>6\)). We refer to the dichotomized GHQ-sum score as psychological distress and the dichotomized GHQ-b subscale score as anxiety and sleep issues. We emphasize that this binarization was chosen for assessing stressor effects in a presence/absence sense (independent of frequency or magnitude) and it is not a requirement of the MiniTransformer architecture, which can, in principle, incorporate continuous-valued variables and missingness-aware representations. The first dataset \(\mathcal{D}_1\) includes the following items: nightmares (dh\_10), sleep problems (dh\_35), paperwork (dh\_37), housekeeping (dh\_38), noise (dh\_45), long work hours (dh\_53), financial problems (le\_8), arguments with a partner (le\_17), serious illness (le\_22), and a summary measure for Anxiety and Sleep issues (subscale GHQ-b). The second dataset \(\mathcal{D}_2\) includes commute to work/school (dh\_11), unwanted visit (dh\_31), paperwork (dh\_37), housekeeping (dh\_38), bad weather (dh\_42), traffic (dh\_46), lost job (le\_1), breakup (le\_16), arguments with a partner (le\_17), and a summary measure for psychological distress (GHQ-sum). After selecting the variables of interest for these two datasets, we included only individuals with no missing follow-ups or items in both datasets, purely for simplicity in this real-data demonstration and to keep the showcase focused on the core method. In general, missing follow-ups or items can be handled using standard transformer mechanisms such as attention masking. The first dataset comprises \(|\mathcal{D}_1| = 882\) individuals, and the second includes \(|\mathcal{D}_2| = 878\) individuals. In both datasets, individuals have sequences of varying lengths, with a median of 13, a minimum of 3, and a maximum of 20 observations. Note that in autoregressive longitudinal modeling, each observation is conditioned on an individual’s prior history, and repeated measurements within a person are strongly dependent. Thus, many follow-ups do not translate into a proportional increase in independent information, i.e., the effective size is primarily governed by the number of individuals and trajectory diversity, placing LORA in a small-data setting.

We compared the MiniTransformer approach to three other approaches. The first and the second are the same as in the simulation study, i.e. an averaging approach and \(p\) regression models, each trained to predict a single variable from all variables at the previous time point. The third approach just carries the last observed values forward as a prediction for the next time point. 

For the MiniTransformer, we used \(H = 8 \) heads and \(C = 8\) cumulants, and estimated parameters by batched stochastic gradient descent, each batch comprising sequences from two individuals, with a learning rate of \(\eta = 0.001 \) and trained the model for 150 epochs. 

Prediction was assessed across 10 cross-validation folds. Table \ref{tab:real_data_comparison_with_baselines} presents the results. The MiniTransformer approach consistently achieves the lowest prediction error, both when considering the target variable or the average error across all variables.

\begin{table}[tb]
    \centering
    \footnotesize
    \caption{Prediction performance of the MiniTransformer for two resilience datasets, each with $p=10$, compared to a simple averaging approach, regression models, and a carry forward approach. Mean squared error is considered across all variables (MSE) and for a specific target variable (MSE$_{\mathrm{tar}}$), together with standard deviations. Smallest values are indicated by boldface.}
    \label{tab:real_data_comparison_with_baselines}
    \begin{tabular}{p{3.0cm}p{1.5cm}cc}
        \toprule
        \textbf{Approach} & \textbf{Metric} & Dataset 1 & Dataset 2 \\
        \midrule
        Average  & MSE  & $0.170 \pm 0.008$    & $0.149 \pm 0.006$    \\
                              & MSE$_{\mathrm{tar}}$ & $0.229 \pm 0.027$    & $0.208 \pm 0.018$    \\
        \addlinespace
        Carry forward   & MSE  & $0.220 \pm 0.013$    & $0.212 \pm 0.020$    \\
                              & MSE$_{\mathrm{tar}}$ & $0.304 \pm 0.013$    & $0.298 \pm 0.020$    \\
        \addlinespace
        Regression        & MSE  & $0.147 \pm 0.009$    & $0.135 \pm 0.008$    \\
                              & MSE$_{\mathrm{tar}}$ & $0.199 \pm 0.024$    & $0.190 \pm 0.023$    \\
        \addlinespace
        MiniTransformer   & MSE  & $\mathbf{0.139 \pm 0.011}$ & $\mathbf{0.127 \pm 0.009}$ \\
                              & MSE$_{\mathrm{tar}}$ & $\mathbf{0.184 \pm 0.021}$ & $\mathbf{0.182 \pm 0.025}$ \\
        \bottomrule
    \end{tabular}
\end{table}

For the statistical testing approach, we used sampling with  \(V = 8\) visit samples. For the first dataset, we consider anxiety and sleep issues (subscale GHQ-b) as the target, and for the second dataset, psychological distress (GHQ-sum). 
The results are presented in Table~\ref{tab:real_data_statistical_testing}. 

In the first dataset, housekeeping (dh\_38) had the lowest p-value, followed by long work hours (dh\_53), suggesting that these contexts may play an important role in influencing how other variables in subsequent observations predict the target. In the second dataset, housekeeping (dh\_38) was identified as the most important context variable, followed by paperwork (dh\_37), traffic (dh\_46), and commute to work/school (dh\_11). At first, these results may seem surprising because of the high emphasis on the housekeeping stressor. Still, upon closer examination of the data, we observe that in the second dataset, housekeeping is a very common reported stressor, i.e., it is reported in 94\% of the time points. On the other hand, it is a well-known finding that cessation of housekeeping often leads to a cluttered environment that may trigger or worsen mental distress, whereas keeping up with basic cleaning can support mental well-being \citep{rosterDarkSideHome2016}. 

To see if the results from our approach confirm this pattern, we performed a check for the second dataset, where we detected housekeeping as a significant context for predicting psychological distress (GHQ\_sum). For this, we assumed an imaginary individual, who had reported all events at the first time point (\(x_{t_1, j} = 1, \forall j \in \{1,\ldots, p+1\}\)), and then we defined different scenarios for the second time point and compared the predicted value. As the first scenario, we assumed that the same pattern is reported at the second time point, i.e., \(\mathbf{x}_{t_1} = \mathbf{x}_{t_2}\). Here the model predicts the target as \(\hat{y}_{t_3, 10} = 0.85\). In the second scenario, we kept all the stressors except for housekeeping (\(x_{t_2, 5} = 0\)). Using this, the model predicted \(\hat{y}_{t_3, 11} = 0.89\). In the third scenario, we switched all the stressors at time-point \(t_2\) off except for housekeeping (\(x_{t_2, 5} = 1\)), and the model predicted \(\hat{y}_{t_3, 11} = 0.38\). As a final scenario, we defined \(x_{t_2, j} = 0, \forall j \in \{1, \ldots, p+1\}\), and here the model prediction changed to \(\hat{y}_{t_3, 11} = 0.42\). With these examples, we can see that in both changes from scenario one to two and scenario three to four, stopping housekeeping leads to an increase in the predicted value for psychological distress (GHQ\_sum).

\begin{table}[tbp]
\centering
\caption{Results of the permutation testing approach for the two resilience application datasets. Average p-values are reported together with standard deviations. Variables are ranked in ascending order of p-values, with rank 1 corresponding to the most important variable.}
\label{tab:real_data_statistical_testing}
\renewcommand{\arraystretch}{1.2}
\footnotesize
\begin{tabular}{ccccc}
\toprule
\multirow{2}{*}{Rank} 
  & \multicolumn{2}{c}{Dataset 1} 
  & \multicolumn{2}{c}{Dataset 2} 
  \\
\cmidrule(lr){2-3} \cmidrule(lr){4-5}
& Variable & pval 
& Variable & pval 
 \\
\midrule
$1$  & dh\_38 & \(0.0131 \pm 0.02\)     & \textbf{dh\_38} & \(\mathbf{0.0001 \pm 0.00}\)\\
$2$  & dh\_53 & \(0.0720 \pm 0.08\)     & dh\_37          & \(0.0346 \pm 0.05\)         \\
$3$  & le\_8  & \(0.1321 \pm 0.16\)     & dh\_46          & \(0.0850 \pm 0.05\)         \\
$4$  & dh\_37 & \(0.1559 \pm 0.08\)     & dh\_11          & \(0.0996 \pm 0.03\)         \\
$5$  & dh\_45 & \(0.2240 \pm 0.11\)     & le\_17          & \(0.3613 \pm 0.10\)         \\
$6$  & dh\_10 & \(0.2631 \pm 0.12\)     & le\_16          & \(0.3631 \pm 0.12\)         \\
$7$  & ghq\_b\_sum & \(0.3326 \pm 0.43\) & le\_1           & \(0.3813 \pm 0.02\)         \\
$8$  & le\_22 & \(0.3732 \pm 0.08\)     & dh\_31          & \(0.4589 \pm 0.11\)         \\
$9$  & dh\_35 & \(0.3872 \pm 0.17\)     & ghq\_sum        & \(0.6965 \pm 0.42\)         \\
$10$ & le\_17 & \(0.6141 \pm 0.15\)     & dh\_42          & \(0.7957 \pm 0.13\)         \\
\bottomrule
\end{tabular}
\end{table}

\begin{figure}[htbp]
    \centering
        \includegraphics[width=8cm]{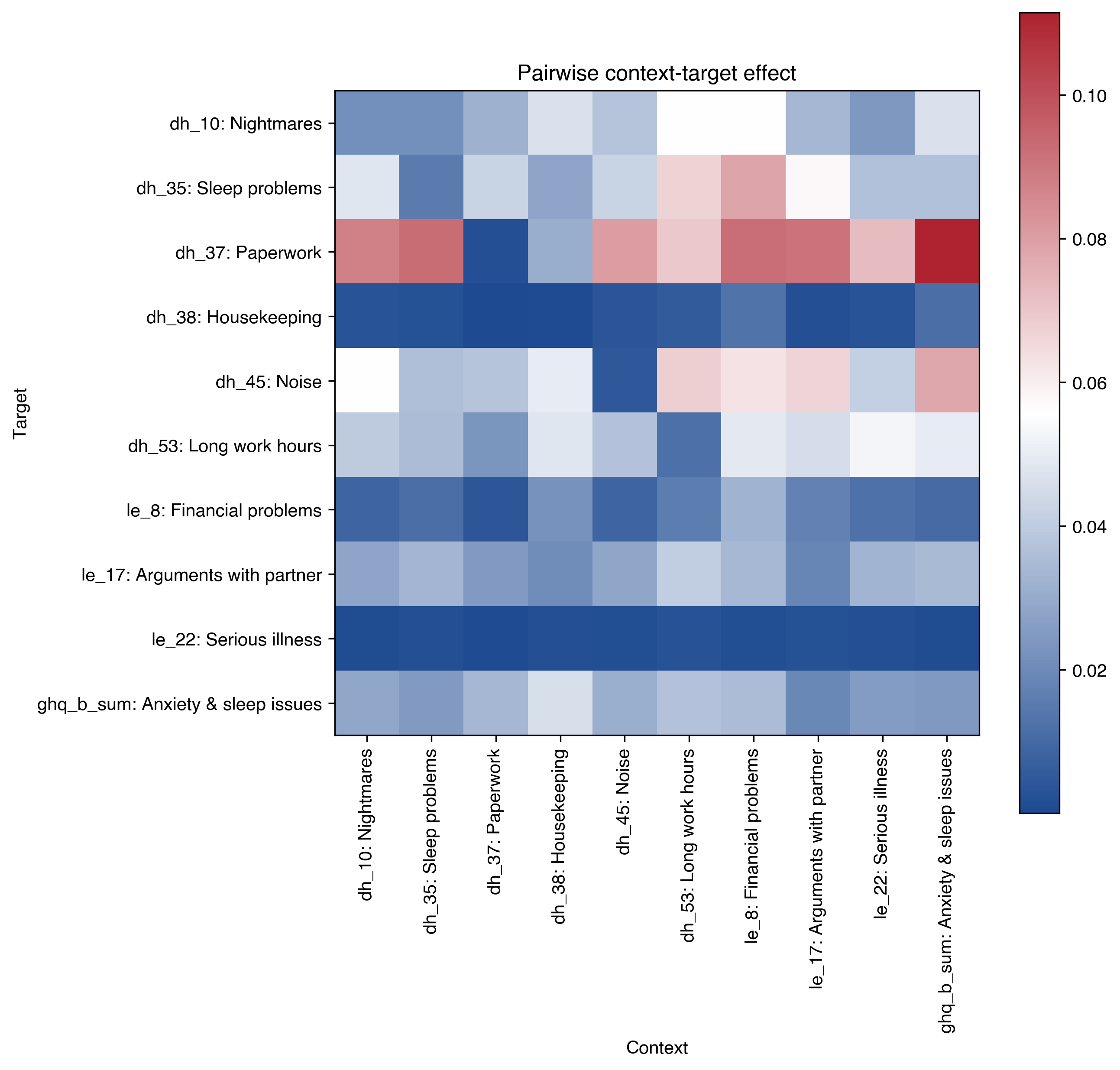}
    \hfill
        \includegraphics[width=8cm]{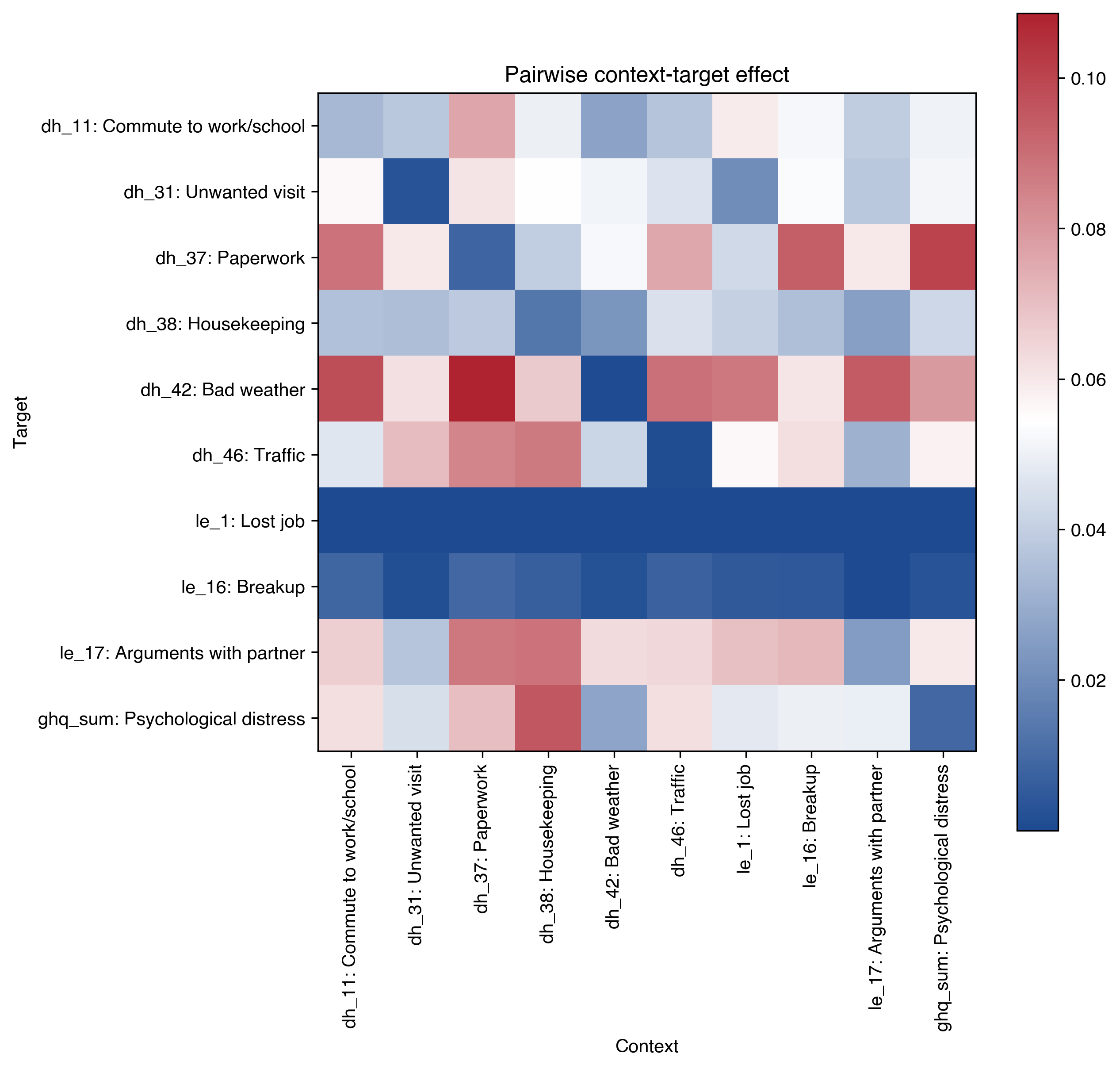}
    \caption{Context-target effects for resilience dataset 1 (left) and dataset 2 (right).
    Dark blue indicates smaller values of the visualized test statistic \(s^{(r)}_j\) (i.e., the context 
    has a smaller effect), while dark red indicates larger values.}
    \label{fig:combined_side_by_side}
\end{figure}

The heatmaps in Figure \ref{fig:combined_side_by_side} illustrated the matrix $\mathbf{S}$ of test statistics $s_j^{(r)}$ for both datasets, to highlight context-target effects. No specific context seems to be important when predicting serious life events, e.g., serious illness (le\_22) in the first dataset and lost job (le\_1) and breakup (le\_16) in the second dataset. This is in line with the expectation that serious life events typically are not triggered by daily hassles. 
Instead, mental health status-related events reported by the GHQ questionnaire may have a greater influence on minor daily hassles, i.e., individuals are more likely to report daily hassles, such as paperwork. This also complies with the  Mood-congruent memory bias, which is a well-known challenge in mental health research that an individual’s current mental state can influence how they perceive and report stress, and suggests that people experiencing depression or anxiety are more likely to recall and emphasize negative or difficult life events \citep{losiakStressfulLifeEvents2019}. 

\section{Discussion}\label{sec:discussion}

We introduced the MiniTransformer approach, a simplified decoder-only transformer architecture optimized for longitudinal data settings with small sample sizes and a few time points. This was complemented by a permutation testing approach that assesses whether a given context significantly modifies the effect on predictions over time. Due to the flexibility of the patterns that can be picked up by the attention mechanism of a transformer, this statistical testing approach can uncover context effects that reflect complex temporal patterns. Specifically, it quantifies through which variables past observations influence predictions based on the most recent observations, and how these relationships evolve throughout a sequence. In addition, a visualization based on corresponding test statistics can highlight context–target associations across time, offering an interpretable view of temporal dependencies.

We evaluated and illustrated these approaches using a simulation design with different sample sizes and a real data application based on two subsets of a dataset collected in a resilience study. The results suggested that the MiniTransformer approach can uncover patterns even with a very limited number of sequences. Overfitting did not seem to be a strong concern, despite still having to estimate a considerable number of parameters.

Naturally, there could have been alternative ways for simplifying the standard transformer architecture, as it has many components that could be adapted. We did not attempt an evaluation of different kinds of adaptations because of the large number of variants that could potentially be considered. Instead, we chose an approach where we used a rather simple statistical model class, VAR models, as a starting point, and added in only a minimal number of core transformer ideas. This subsequently also enabled an interpretable test statistic for assessing context effects of variables.

A limitation of the proposed statistical testing approach is the limited sample size for the empirical distribution in settings with a small number of variables. With a larger number of variables, computational feasibility is a concern. Generating a full null distribution requires a large number of permutations. When having to switch to a sampling-based approach due to computational cost,  the statistical power of the test may be affected, particularly in capturing subtle differences between contexts. This limitation may lead to an increased risk of false negatives, where potentially informative contexts are overlooked due to insufficient permutation-based significance estimation. 

It is worth noting that, although the current implementation focuses on binary variables, the underlying framework can be readily extended to continuous variables. In such cases, contexts could be defined through discretization or binning schemes to capture graded effects. While binary variables simplify interpretation, extending the approach to continuous contexts also would enable a more fine-grained assessment of how varying levels of a variable influence predictions. This generalization would further enhance the applicability of the MiniTransformer approach to diverse clinical and longitudinal datasets.

In summary, the proposed MiniTransformer approach more generally demonstrates that transformers can indeed be adapted for small longitudinal datasets with few time points, and then not only can achieve reasonable prediction performance but also can offer interpretability. Therefore, it might be promising to consider simplified transformer architectures also for other settings with sequential structure in cohort data.

\section*{Author contributions}

HB and KF developed the MiniTransformer architecture and the statistical testing approach. KF performed the simulation study and the evaluation on the real data. MHa contributed to the mathematical formulation of the method. MW and MHe provided critical input on the statistical methodology and its interpretation. The LORA consortium co-authors (KA, CS, BK, OT, KL, MP, AR, and RK) contributed to data provision as part of the LORA study. All authors reviewed and approved the manuscript.

\section*{Acknowledgments}
The work of Maren Hackenberg, Moritz Hess, and Harald Binder was funded by the Deutsche Forschungsgemeinschaft (DFG, German Research Foundation) – Project-ID 499552394 – SFB 1597.
This project has received funding from the European Union’s Horizon 2020 research and innovation program under Grant Agreement number 777084 (DynaMORE project) and from Deutsche Forschungsgemeinschaft (CRC 1193, subproject Z03). This work is part of the LOEWE Center “DYNAMIC”, funded by the HMWK Hessen.

\section*{Conflict of interest}
The authors declare no potential conflict of interest.

\bibliographystyle{unsrt}  
\bibliography{references}

\end{document}